\documentclass[conference]{IEEEtran}
\IEEEoverridecommandlockouts
\usepackage[normalem]{ulem}
\usepackage{cite}
\usepackage{amsmath,amssymb,amsfonts}
\usepackage{dsfont}
\usepackage{textcomp}
\usepackage{graphicx}
\usepackage[table,xcdraw]{xcolor}
\newcommand{\cmmnt}[1]{}
\def\BibTeX{{\rm B\kern-.05em{\sc i\kern-.025em b}\kern-.08em
    T\kern-.1667em\lower.7ex\hbox{E}\kern-.125emX}}
\begin{document}

\title{AIIPot: Adaptive Intelligent-Interaction  Honeypot for IoT Devices\\
}

\author{\IEEEauthorblockN{Volviane Saphir Mfogo }
\IEEEauthorblockA{
\textit{Department of Mathematics \& Computer Science} \\
\textit{University of Dschang}\\
BP 96 Dschang, Cameroon \\
smfogo@aimsammi.org }
\and
\IEEEauthorblockN{Alain Zemkoho }
\IEEEauthorblockA{
 \textit{School of Mathematical Sciences} \\
\textit{University of Southampton}\\
SO17 1BJ Southampton, UK \\
a.b.zemkoho@soton.ac.uk }
\and
\IEEEauthorblockN{Laurent Njilla }
\IEEEauthorblockA{
\textit{Information Assurance Branch}\\
\textit{Air Force Research Laboratory}\\
 Rome, NY, USA \\
laurent.njilla@us.af.mil}
\and
\IEEEauthorblockN{\qquad\quad  ${}$ }
\IEEEauthorblockA{
\textit{} \\
\textit{}\\
 \\
}
\and 
\IEEEauthorblockN{Marcellin Nkenlifack}
\IEEEauthorblockA{
\textit{Department of Mathematics \& Computer Science} \\
\textit{University of Dschang}\\
BP 96 Dschang, Cameroon \\
marcellin.nkenlifack@gmail.com}
\and
\IEEEauthorblockN{Charles Kamhoua}
\IEEEauthorblockA{
\textit{Combat Capabilities Development Command}\\
\textit{Army Research Laboratory}\\
 Adelphi, MD, USA \\
charles.a.kamhoua.civ@army.mil}}

\maketitle

\begin{abstract}
The proliferation of the Internet of Things (IoT) has raised concerns about the security of connected devices. There is a need to develop suitable and cost-efficient methods to identify vulnerabilities in IoT devices in order to address them before attackers seize opportunities to compromise them. The deception technique is a prominent approach to improving the security posture of IoT systems. Honeypot is a popular deception technique that mimics interaction in real fashion and encourages unauthorised users (attackers) to launch attacks. Due to the large number and the heterogeneity of IoT devices, manually crafting the low and high-interaction honeypots is not affordable. This has forced researchers to seek innovative ways to build honeypots for IoT devices. In this paper, we propose a honeypot for IoT devices that uses machine learning techniques to learn and interact with attackers automatically. The evaluation of the proposed model indicates that our system can improve the session length with attackers and capture more attacks on the IoT network. 
\end{abstract}

\begin{IEEEkeywords}
Honeypot, Internet of Things (IoT) Devices, Machine Learning, Reinforcement Learning.
\end{IEEEkeywords}

\section{Introduction}
  The Internet of Things (IoT) has captured the interest of significant service providers, enterprises, and industries in recent years, including Healthcare, Smart Homes, Autonomous Vehicles, Digital Agriculture, and many others. IoT has transformed the way we live and work by allowing physical devices to communicate with one another and with us via the internet. However, with this increased connectivity comes the need for effective communication and security measures to ensure the safety and privacy of sensitive data transmitted through these systems. 
Unlike traditional computers, IoT devices typically have network interfaces that allow interaction between the physical and virtual worlds. However, they suffer from various vulnerabilities such as weak or hard-coded passwords. Many passwords are easy to guess, publicly available, or cannot be changed, putting them at risk of being compromised easily. Deception technology is a useful approach to improving the security posture of IoT systems. Honeypot is one of the  deception methods to discover vulnerabilities that is commonly used by security practitioners. In general, a  honeypot mimics interaction in real fashion and encourages unauthorised users (attackers) to launch attacks.

\begin{figure}[htbp]
\includegraphics[width=0.5\textwidth]{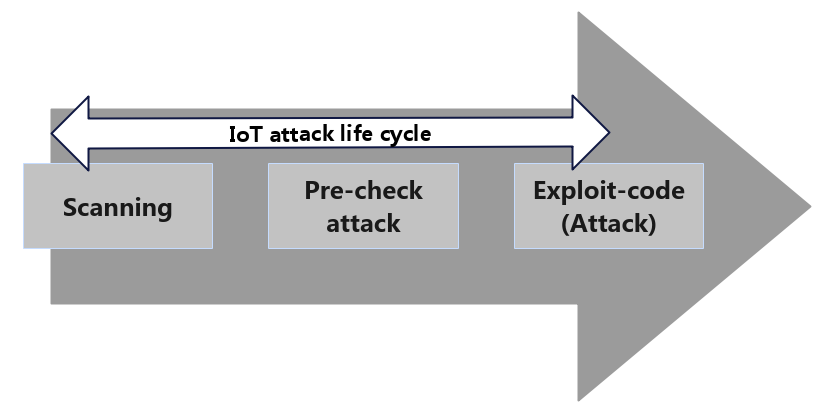}
\caption{Life cycle of IoT attack.}
\label{fig:iotattacklifcycle}
\end{figure}

However, IoT vulnerabilities are usually highly dependent on the nature of the device, firmware version, or even the vendor. This leads to the fact that after scanning the network, attackers observe the network vulnerabilities (like open ports), and tend to perform several checks on the remote target device to gather more information about the specific device before launching the code of attack (exploit-code). This phase is called the pre-check attack step. Figure \ref{fig:iotattacklifcycle} summarises the life cycle of an IoT attack. So a honeypot with a limited level of interaction is not enough to pass the check and will fail to capture the real attack.
Our goal is to build a honeypot that is able to interact with attackers during the phase of checking to observe attacks targeting IoT devices effectively. This is to get access to the exploit code of the attacker. In order to achieve our goal, honeypots must be able to pass the pre-check step of the attacker. Due to the large number and the heterogeneity of IoT devices, manually crafting the low and high-interaction honeypots are not affordable. However, some device checks are simple, such as verifying that the target device’s response is correct. Thus, a honeypot that returns a correct response to the received request may bypass their checks. Other ones can be more complex and require more steps before the exploit code is launched.

In this paper, we proposed a honeypot for IoT devices based on a machine learning concept that we label as \textit{adaptive intelligence-interaction: AIIPot}. AIIPot is a chatbot that is based on a transformer and reinforcement learning models. The chatbot is trained on a dataset of requests/responses in order to choose a response with a high probability to be expected by an attacker to a specific request at the early stage of the attack. However, our honeypot uses the reinforcement learning concept to model the future direction of the conversation with the attacker. With this, AIIPot can respond appropriately to requests, with no need to learn enough to interact with attackers continuously. AIIPot adapts its interaction with the attacker using the fact that all new requests arriving at the system are broadcast to the IoT device network, which gives a high expected response to the attacker. This leads to the collection of more datasets. Then, the AIIPot chatbot model chooses the best response. With these methods, our honeypots return expected responses to the attacker even if initially the request is not present on the database, therefore increasing the likelihood that the attacker will launch the attack on the honeypot believing that it is an IoT device. This allows us to extend the session length, detect and monitor attacks, collect information about the tactics and techniques used by attackers, and ultimately help organizations improve their security defenses.

To the best of our knowledge, this paper is the first to build a honeypot for IoT devices based on a transformer chatbot, which uses reinforcement learning to model the future direction of the conversation.
In summary, the contributions of this work are as follows:
\begin{enumerate}

    \item A honeypot based on a transformer model is proposed to capture vulnerabilities on IoT devices.
    
    \item  Reinforcement learning concepts are used to model the future direction of the interaction between the honeypot and the attacker.
    
    \item A novel technique to collect a new dataset of the interaction between attackers and IoT devices is proposed.
\end{enumerate}

The rest of the paper is structured as follows: Section \ref{sectionBRW} provides the background and related works on IoT honeypots, and Section \ref{sectionAIIPot} presents our approach. Next, we present the evaluation of the proposed honeypot in Section \ref{sectionEval} and conclude in Section \ref{sectionConclu} while discussing possible ideas for future research.

\section{Background and Related Works}\label{sectionBRW}

\subsection{Internet of Things}

The IoT is the vast network of connected physical objects (i.e., things) that exchange data with other devices and systems via the internet. It's characterized by heterogeneous identifiable networking objects (sensors or actuators) advertising their services to assemble semantic-rich applications. The heterogeneity of IoT devices and networks is mainly caused by various manufacturers and (communication) protocols. As a result, they suffer from various vulnerabilities such as weak or hard-coded passwords making them easy to be compromised. 

\subsection{Machine learning for Cybersecurity}
Machine Learning (ML) is an area of artificial intelligence and computer science that uses data and algorithms to mimic how humans learn. It's used in research to create security systems like intrusion detection systems (IDS). Deep Learning is a subfield of machine learning that improves fields such as Natural Language Processing (NLP). Chatbots are NLP apps that deliver automatic responses to consumer enquiries. They can be used to create security mechanisms for IoT cyber deception.

\subsection{Honeypot for Cybersecurity}
Deception falls into six different categories: perturbation, obfuscation, moving target defence, mixing, honey-x, and attacker engagement\cite{b2}. A honeypot is a variant of honey-x and is mainly used to deceive attackers from their actual target or to collect information regarding attack patterns. Honeypot is a technology used to capture attacks on IoT devices. Mainly two categories of honeypots exist: \textit {high} and \textit{low} interaction. In between low and high, hybrid/medium interaction also exists. In addition to these categories, this paper describes another interaction level, intelligent-interaction honeypots based on machine learning. \cmmnt{This survey paper \cite{b3} revisited all of the honeypot research projects since 2002.}

\subsection{Related Works}
\subsubsection{Low-Interaction Honeypots}
Low-interaction honeypots are just emulated services and give the attacker a very limited level of interaction, such as a popular one called honeyd \cite{b4}. IoT low-interaction honeypot emulates a single protocol, and/or emulates a specific device such as U-PoT \cite{b5} and ThingPot \cite{b6}, respectively. 
Low-interaction honeypots support only some functions of the system, not the entire system; therefore, their fixed behaviour when receiving a non-emulated request makes them limited and easily detectable by the attacker.

\subsubsection{High-Interaction Honeypots}

High-interaction honeypots are fully-fledged operating systems and use real systems for attackers to interact with. They collect advanced information on cyber attacks by providing systems over which attackers have complete control. SIPHON \cite{b10} is an example of a high-interaction IoT honeypot  which deploys a physical device as a honeypot. 
The problems of high interaction are their complexity and the needed time for deployment and maintenance.
So as the number of honeypots increases, scalability decreases. The reason is that physical devices and virtual machines consume computational resources.



\subsubsection{Intelligent-Interaction Honeypots}
The concept of intelligent-interaction honeypots is an interaction with attackers that maximize the likelihood of catching the attacks instead of accurately emulating the behaviour of a specific service or device as in high, low, or hybrid interaction. However, there are few research models that stand out as the most versatile as they emulate full devices and are self-adaptive: IoTCandyJar \cite{b7}, Chameleon \cite{b8}, and FirmPot \cite{b9}. 
Intelligent-interaction honeypot was introduced for the first time in 2017 on IoTCandyJar \cite{b7}.  

IoTCandyJar \cite{b7} proposed an intelligent-interaction IoT honeypot that can emulate all kinds of IoT devices. For a specific request, IoTCandyJar selects the response the attacker expects from many responses of IoT devices collected by internet scanning. If the selected response is the expected one, the attackers assume that the honeypot is their target device and send an exploit code. They used a Markov Decision Process model (MDP) to learn from scratch what response an attacker expects. Thus, the problem is that their honeypots take some time until they can respond appropriately to requests. As a result, it took two weeks for the model to learn enough to interact with attackers continuously. Another problem is that the responses collected from different IoT devices by internet scanning  are not 100\% sure that they are from IoT devices as they could be from some honeypot on the internet as well. 


FirmPot \cite{b9} proposed a framework that automatically generates intelligent-interaction honeypots using firmware. This framework collects web interactions by emulating firmware launched on a docker container and learns the correspondence of requests and responses by machine learning. The generated honeypots return the best from the emulated responses to the received request.
\begin{itemize}
    \item  Firstly, the honeypots generated by this framework do not capture any advanced attacks such as configuration changes. This is due to the fact that (1) either there is the issue of the honeypot observation location and periods; (2) or the Seq2Seq learning model used by the authors does not converge or the learning model is unable to interact; (3) or in the worse case, the generated honeypots have been detected by attackers, which is a general problem with most existing honeypots.
    \item Secondly, during the scanning process, the responses collected could be from fake web applications (same as a honeypot).
    \item Finally, this approach highly depends on the firmware images; so the vulnerability can be discovered when generating a honeypot depending on the vulnerability of the target vendor.
\end{itemize}
To overcome all the above limits, we proposed an effective honeypot based on a machine learning concept that responds correctly to attackers at an early stage of the attack. The proposed honeypot for IoT devices uses machine learning techniques to learn and interact with attackers automatically.

\section{AIIPot}\label{sectionAIIPot}
In this section, we \cmmnt{clearly} describe in detail our approach to building a honeypot for IoT devices. 

\subsection{Overview}

\begin{figure}[htbp]
\centerline{\includegraphics[width=0.5\textwidth]{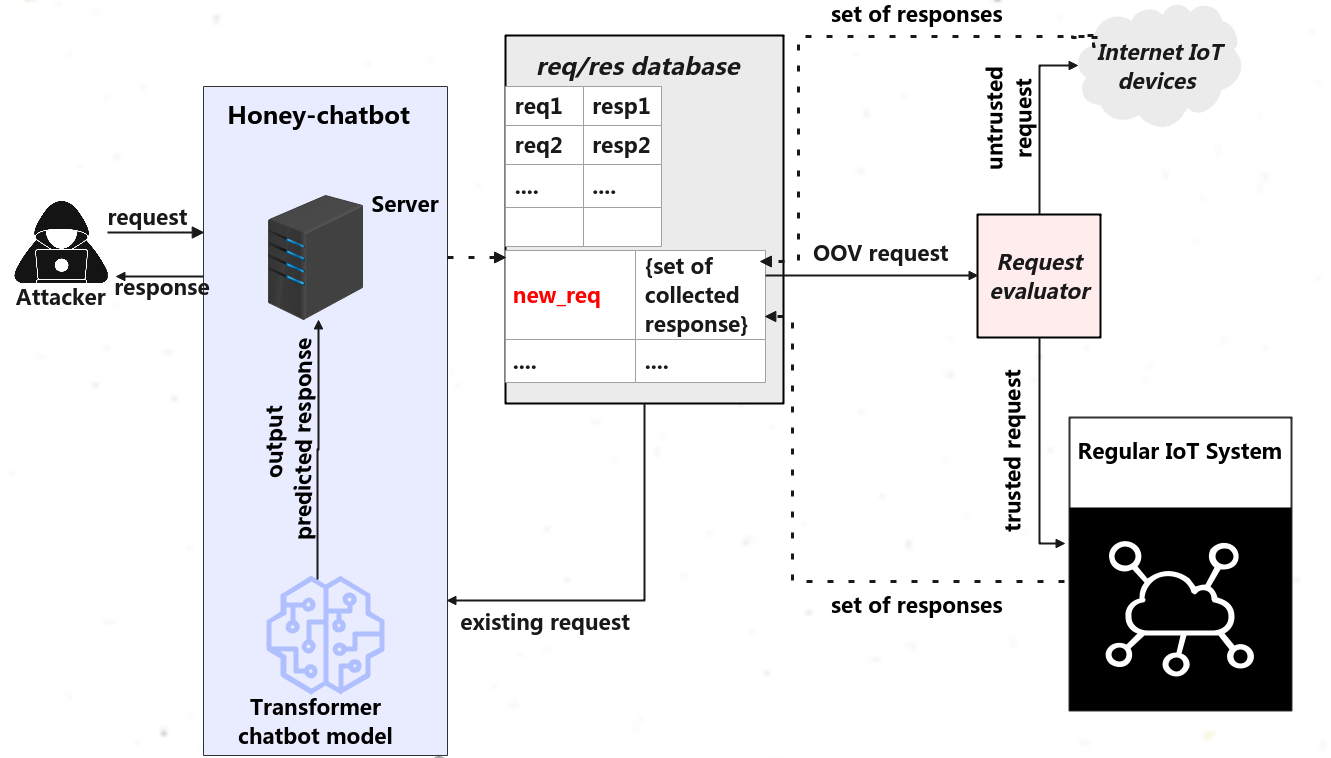}}
\caption{High-level overview of AIIPot.}
\label{fig:aiipot}
\end{figure}

The high-level overview of our approach is shown in figure \ref{fig:aiipot}. Our framework has three components: the \textit{honey-chatbot, req/res database}, and \textit{request evaluator}. The \textit{honey-chatbot} module is used to reply to the request of an attacker on the honeypot.
The dataset is saved on the req/res database, which is a database of possible requests that an attacker can send to an IoT device and corresponding responses that the IoT device can reply to the attacker. If the request is not part of the req/res database, this is considered as an \textit{out-of-vocabulary} (OOV) request, and then we saved that request to a new line of our \textit{req/res database} and send the request to the \textit{evaluator} module for security evaluation. The evaluator module evaluates the request's trust and either broadcasts it to all the devices of the local network in the case where the request is trustable\footnote{The request doesn't contain any exploit code } or sends it to the IoT device on the internet for the untrusted ones.

\subsection{Honey-Chatbot}
The module Honey-Chatbot corresponds to a server where the response model selection is deployed. With the help of the req/res database, our honeypot is enabled to reply with a valid response to the client based on the received request instead of responding to the fixed one. In this section, we discuss how to leverage the transformer and the MDP model to optimize the response selection with the maximal possibility to capture attacks.

\subsubsection{Honey-Chatbot Overview}
For each individual request, the req/res database module could contain at least hundreds or even thousands of responses. All of them are valid responses, but only a few of them are the correct and expected ones. This is because, for a given request, various IoT devices can respond to it under their own logic to process it and generate the response. The most straightforward example is the request to access the root path of their web service: some devices may reply by the login portal page, others may redirect it to another resource, and the rest may respond with different types of error pages. Therefore, all of the responses in the req/res database are potential candidates as the response to the attacker, but the challenge is to find the one that is expected by the attacker.

\textbf{Our approach}: the idea behind our approach is to first train a transformer-based architecture (BERT) model on the dataset present in the req/res database and record the possible response candidate that is likely to be expected by an attacker, then choose the expected one with the high probability with an MDP model and record the next move from the attacker's side. We assume if we happen to select the correct one, attackers will believe our honeypot is the vulnerable target IoT device and continue to extend the session length and eventually send the malicious payload. Every incoming request to the honeypot will be forwarded to this module, and the selected response will be returned to the client. The core part of the module is the selection engine, which passes the request into the BERT model and fetches the potential responses list from the model output of all the transactions for the MDP model selection. In MDP selection mode, it first locates the state in the graph from the normalized request and is followed by the model to select the best one.


\subsubsection{Model Formulation}
We discuss how we formulate the response selection problem with the transformer-based architecture (BERT) and MDP model. We assume whether the client continues the session or performs the attack is simply determined by the response to the previous request. This is a reasonable assumption based on our best knowledge of the existing malware samples.

\textbf{Transformer-based architecture (BERT)}: Bidirectional Encoder Representations from Transformers (BERT) \cite{b20} belongs to the
family of the so-called transfer learning methods,
where a model is first pre-trained on general tasks
and then fine-tuned the final target tasks. Transfer learning has been shown beneficial in many different tasks. BERT is a very deep model that is pre-trained over large corpora of raw texts and then fine-tuned on target annotated data. The building block of BERT is the Transformer \cite{b18}, an attention-based mechanism that learns contextual relations between words (or sub-words, i.e., word pieces) in a text. BERT provides contextualized embeddings of the words composing a sentence as well as a sentence embedding capturing sentence-level semantics: the pre-training of BERT is designed to capture such information by relying on very large corpora. These can be used in input to further layers to solve sentence classification: this is achieved by adding task-specific layers and by fine-tuning the entire architecture on annotated data.
Initially, we created a dictionary that maps the words of the request to numerical values. Next, a word embedding is trained from all the numerical requests and responses vector $x_t$ and $y_t$, respectively. The vectorized requests and responses are fed into the BERT model. We extend BERT by using task-specific layers, as in the usual BERT fine-tuning. This outputs $\pi_{\theta}(y_t|x_t)$, the probability of choosing the response $y_t$ given the input request $x_t$. By choosing the response corresponding to $\pi_{\theta}(y_t|x_t)$ there is no guarantee that this predicted response will push the attacker to send a new request. So in order to model the future of the interaction encouraging the attacker each time to send a new request until the exploit code is sent, we used an MDP to model the next move of the attacker.

    
    

\textbf{Markov decision process (MDP)}: MDP is an extension of the standard (unhidden) Markov model. It is a model for sequential decision-making when outcomes are uncertain, such as computing a policy of actions that maximize utility with respect to expected rewards. At each decision epoch, the next state will be determined based on the chosen action through a transition probability function. The mechanism is collectively referred to as reinforcement learning. Reinforcement learning is a mechanism to control and adjust policy when the reward of the current state space is uncertain.


\textbf{Problem formulation}: in the standard reinforcement learning model, an agent (attacker) interacts with its environment (honeypot). This interaction takes the form of the agent sensing the environment and based on input choosing an action to perform in the environment. Every reinforcement learning model learns a mapping from situations to actions by trial-and-error interactions with a dynamic environment. The model consists of multiple variables, including decision epochs ($t$), states ($x,s$), transitions probabilities ($T$), rewards ($r$), actions ($y$), value function ($V$), discount ($\gamma$), and estimation error (e). The basic rule of a reinforcement learning task is the Bellman equation:
\begin{equation} \label{eq0}
V^{*} (x_t ) = r(x_t ) + \gamma V^* (x_{t+1} ).
\end{equation}
The general update policy can be expressed as
\begin{equation} \label{eq00}
\Delta w_t = \max_y~\left[r(x_t , y) + \gamma V (x_{t+1} )\right]- V(x_t ).
\end{equation}
 Our problem is essentially a non-deterministic Markov Decision Process, which means at each state, there exists a transition probability function T to determine the next state. In other words, our learning policy is a probabilistic trade-off between \textit{exploration, reply with responses which have not been used before}, and \textit{exploitation reply with the responses which have known high rewards}. To apply general valuation iteration is impossible to calculate the necessary
integrals without added knowledge or some decision modification. Therefore, we apply Q-learning to solve the problem of having to take the max over a set of integrals. Rather than finding a mapping from states to state values, Q-learning finds a mapping from state/action pairs to values (called Q-values). Instead of having an associated value function, Q-learning makes use of the Q-function. In each state, there is a Q-value associated with each action. The definition of a Q-value is the sum of the reinforcements received when performing the associated action and then following the given policy thereafter.
Therefore, in our problem of using Q-learning, the equivalent of the Bellman equation is formalized as
\begin{equation} \label{eq000}
 Q(x_t , y_t ) = r(x_t , y_t ) + \gamma \max_{y_{t+1}} Q(x_{t+1} , y_{t+1} ),
\end{equation}
and the updated rule of direct Q-learning is formalized as follows where $\alpha$ is the learning rate

\begin{equation} \label{eq1}
\begin{split}
\Delta w_t = \alpha [(r(x_t , y_t ) +& \gamma \max_{y_{t+1}} Q(x_{t+1} , y_{t+1} , w_t )\\
    &- Q(x_t , y_t , w_t )]\dfrac{\partial Q(x_t , y_t , w_t )}{\partial w_t}.
\end{split}
\end{equation}


\textbf{Reward function}: reward function $r : (x_t, y_t ) \longrightarrow r$
assigns some value $r$ to be in the state and action pair $(x_t , y_t )$. The goal of the reward is to define the preference of each pair and maximize the final rewards (optimal policy). In our context, the immediate reward $r(x_t, y_t )$ reflects the progress we have made during the interaction process when we choose response $y_t$ to request $x_t$ and move to the next state $x_{t+1}$. Since the progress can be either negative or positive, the reward function can be negative or positive as well. The heuristics of defining reward is that if the response is the target device type expected by the attacker and the attacker launches the attack by sending the exploit code in the next request, the reward must be positive and huge. On the contrary, if the response is not an expected one (e.g., reflects a not-vulnerable device version), the attacker may stop the attack and end the session. It leads to a dead-end state and causes a negative reward. In other words, we reward the responses that could lead us to the final attack packet and punish the ones that lead to the dead-end session. One of our designs is to assign a reward as a value equal to the length of the final sessions since we believe the longer request sent by the attackers, the higher chance the malicious payload is contained. The standard session is 2, which means after we send our response, there is at least another incoming request from the same IP at the same port. If no further transition is observed, we assign a negative reward for that response. Another alternative reward assignment could be based on whether we receive some known exploits packets or not.

\textbf{State and action}: in our case, the state $x_t$ corresponds to the requests sent by the attacker with all the similar existing requests on the req/res database. The actions are characterised by the output response of the transformer model  such that $\pi(y_t|x_t)\geqslant threshold$. We would like to fix the threshold at $0.5$. This assumption is realistic due to the fact that if the transformer predicts a response with $0.5$ probability that means there is a high chance that such a response belongs to the set of responses of a vulnerable IoT device.

\textbf{Transition probabilities}: the transition probabilities can be described by the transition function $T(s, y, s')$, where $y$ is an action moving taken during the current state $s$, and $s'$ is some new state. More formally, the transition function $T(s, y, s')$ can be described by the formula
\begin{equation}
    P(S_t = s' |S_{t-1} = s, y_t = y) = T(s, y, s').
\end{equation}

To measure the probability of each combination of $(s, a, s')$, we deployed the trained BERT model that returns a response from the candidate set and saved the session information to the session table. After running a period of time, we are able to collect lots of sessions, and we parse each of them to count the occurrence of each combination $(s, a, s')$, which is denoted as $C(s, a, s')$. The transition function $T(s, y, s')$ are defined as follows:
\begin{equation}
    T(s, a, s') = \dfrac{C(s, a, s')}{\displaystyle\sum_{x \in S}^{} (s, a, x)}.
\end{equation}

\textbf{Online Q-learning algorithm}: we apply the online Q-learning algorithm to select the expected response from the candidate response output by the BERT model. Based on the Q-learning model, our learning process starts from receiving a request at the $t_0$ decision epoch. Given the request, we passed it into the trained BERT model to select a set of candidate responses. We adopt the $\epsilon$-greedy policy for action selection. In particular, we consider probabilities output by the BERT model for each candidate response as the initial transaction functions. Using this policy, we can select random action with $\epsilon$ probability or an action with $1-\epsilon$ probability that gives maximum reward in a given state.
Then we start our Q-learning iteration and update the Q-learning table. When we learn to reinforce for one state and action pair, $r(x_t, y_t)$, we first back-propagate and update the Q lookup table. According to this, we can make the adjustment by removing the responses that end with negative rewards and updating the $epsilon$ value. The iteration runs until the model converges. In practice, our model is running online and updated in real time. Therefore, thanks to the trained BERT model our model has a high chance to converge.

\subsection{Req/Res Database and Request Evaluator}
In this section, we describe in detail how we collect the req/res database component of the proposed approach.

The Honey-Chatbot component used the database during the offline training of the transformer model.\footnote{In an offline machine learning model, the weights and parameters of the model are updated while simultaneously attempting to lower the global cost function using the data used to train the model}

\begin{figure}[htbp]
\centerline{\includegraphics[width=0.5\textwidth]{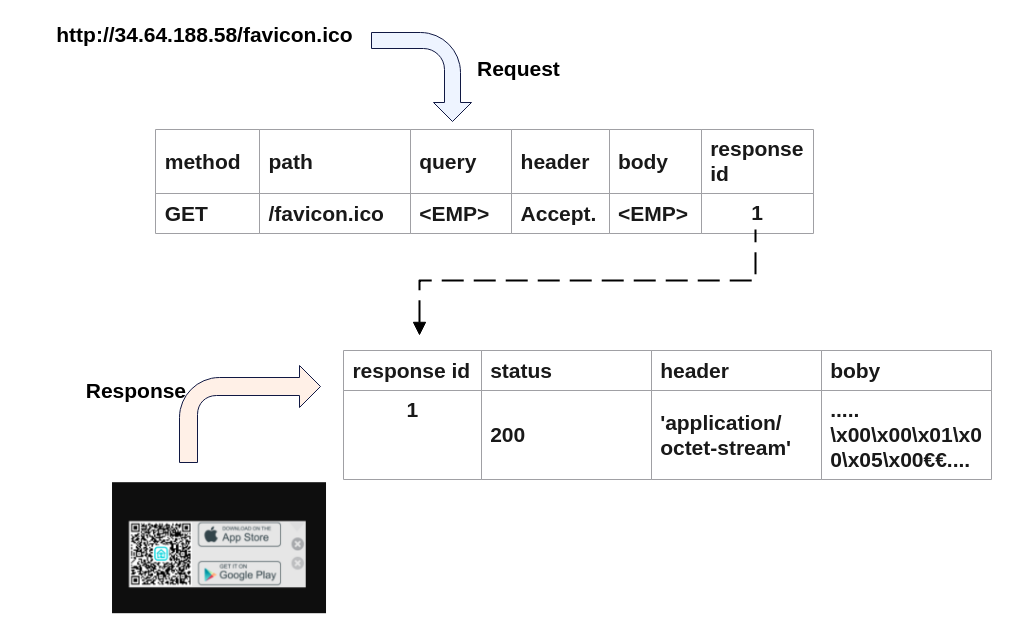}}
    \caption{Req/res database entries. }
    \label{fig:reqres}
\end{figure}

We used the dataset provided by \cite{b9} \cmmnt{as} to form the baseline of our database.
In our database, each entry corresponds to a specific request sent by an attacker and the corresponding response from the IoT device. 
For new requests, we used the \textit{Request Evaluator} to evaluate the trusted request before collecting the response corresponding to that specific request. All new entries request/response(s) are saved on the database as shown in Figure \ref{fig:reqres}.

\begin{figure}[htbp]
\centerline{\includegraphics[width=0.5\textwidth]{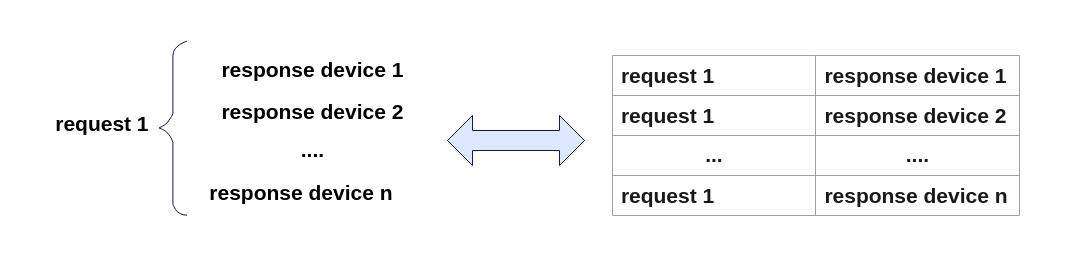}}
    \caption{Request/response multi-entries.}
    \label{fig:databaseentry}
\end{figure}

The request evaluator checks whether the request is an exploit code or not. This module is based on a Super Vector Machine (SVM) classifier. We trained an SVM model on the NSL-KDD dataset \cite{b19} to determine whether a request is an attack or not. Classified an attack is considered an untrusted request and those classified as normal are considered to be a trusted one. If the request is trusted, then we forward it to our IoT local network as shown in figure \ref{fig:aiipot}. Otherwise, the request is forwarded to some IoT devices on the internet that is accessible. We received  many responses from real IoT devices for one request, and we saved them as many entries to the database.
Figure \ref{fig:databaseentry} shows how the new entry is saved on the database.

\section{Evaluation}\label{sectionEval}

\textbf{Datasets}: for the offline training of the transformer model, we used the HTTP protocol dataset. Besides HTTP protocol, a preliminary check happens on even customised IoT protocols. Home Network Administration Protocol (HNAP) is one example. The experimentation was carried out only on the HTTP protocol dataset due to the lack of a dataset on the other existing protocol.  The initial dataset was provided by \cite{b9}. 
The dataset contains about 17,604 entries of requests and the corresponding response. We also use the NSL-KDD dataset \cite{b19} for the request evaluator module.

\textbf{Training schedule of the BERT model}: we use the AdamW optimizer for training for 300 epochs with a cosine decay learning rate schedule. The initial learning rate is set to 0.001 and the batch size is 512.

    \textbf{Evaluation metrics}: we control the performance of our proposed Honey-Chatbot on four metrics; the number of requests captured by the honeypot, the session length, the volume of information sent by an attacker in a single session time, and the number of attack types captured by the honeypot.

    \textbf{Honey-Chatbot deployment}: to evaluate the performance of our proposal, we set up our honeypot on the public server on the Google Cloud Platform (GCP). Due to resources constraint, each method was deployed on the server for $ 20$ days only. Table \ref{tab:my-table} presents the total number of requests that the server received during that period and the total number of IP addresses observed. 
For the sake of comparison to our proposal, we consider the model where the response is chosen randomly among all responses and the honeypots proposed by \cite{b7} and \cite{b9}.

\begin{figure}[htbp]
\centerline{\includegraphics[width=0.5\textwidth]{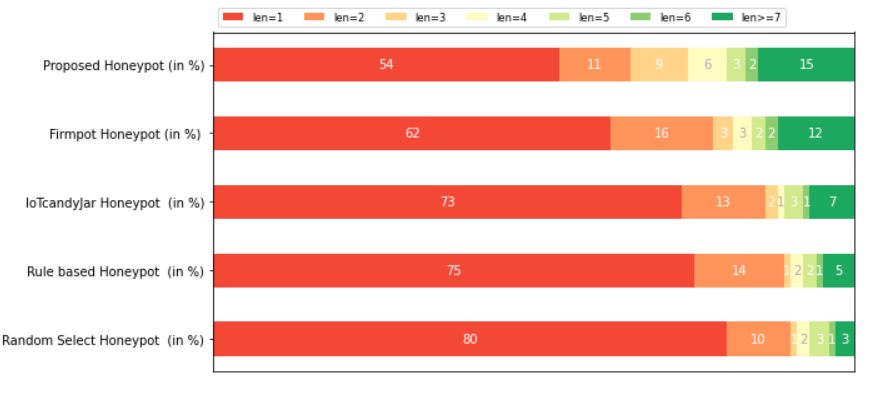}}
    \caption{Comparison of session length with attackers' between honeypots (in percentage). If an attacker sends one request and the honeypot returns one response, and the communication is terminated, the session length is 1. The longer the session length, the more likely an attacker believes the honeypot is a natural IoT device. Hence the session length is considered one of the critical indicators of a honeypot’s deception performance, the effectiveness of the learning process.}
    \label{fig:experiment}
\end{figure}

The proposed Honey-Chatbot in 20 days captures 6,235 requests from 987 different IP addresses. In Figure \ref{fig:experiment} the amount of session length of 1 has been reduced while the session length greater or equal to 7 has increased compared to the existing honeypot \cite{b7} and \cite{b9}. This is because the transformer model helps the MDP model to converge quickly and capture the attention of the attacker with the expected response at each session time.
\begin{table}[htbp]
\caption{Number of requests received by each honeypot on the GCP web server.}
\centering
\begin{tabular}{|c|c|}
\hline
\rowcolor[HTML]{C5D1F6} 
{\color[HTML]{333333} \textbf{Method}} &
  \textbf{\begin{tabular}[c]{@{}c@{}}Total number of requests \\ received / total number of IP addresses\\  observed\end{tabular}} \\ \hline
\cellcolor[HTML]{FFFFFF}\begin{tabular}[c]{@{}c@{}}\textbf{AIIPot} \end{tabular} &
  \cellcolor[HTML]{FFFFFF}6,235/987 \\ \hline
\cellcolor[HTML]{FFFFFF}\begin{tabular}[c]{@{}c@{}}Firmpot \end{tabular} &
  \cellcolor[HTML]{FFFFFF}2,896/687 \\ \hline
\begin{tabular}[c]{@{}c@{}}IoTcandyJar  \end{tabular} &
  2,111/602 \\ \hline
\begin{tabular}[c]{@{}c@{}}Random select honeypot\end{tabular} &
  2,298/345 \\ \hline
\end{tabular}%
\label{tab:my-table}
\end{table}
  
\begin{figure}[htbp]
\centerline{\includegraphics[width=0.5\textwidth]{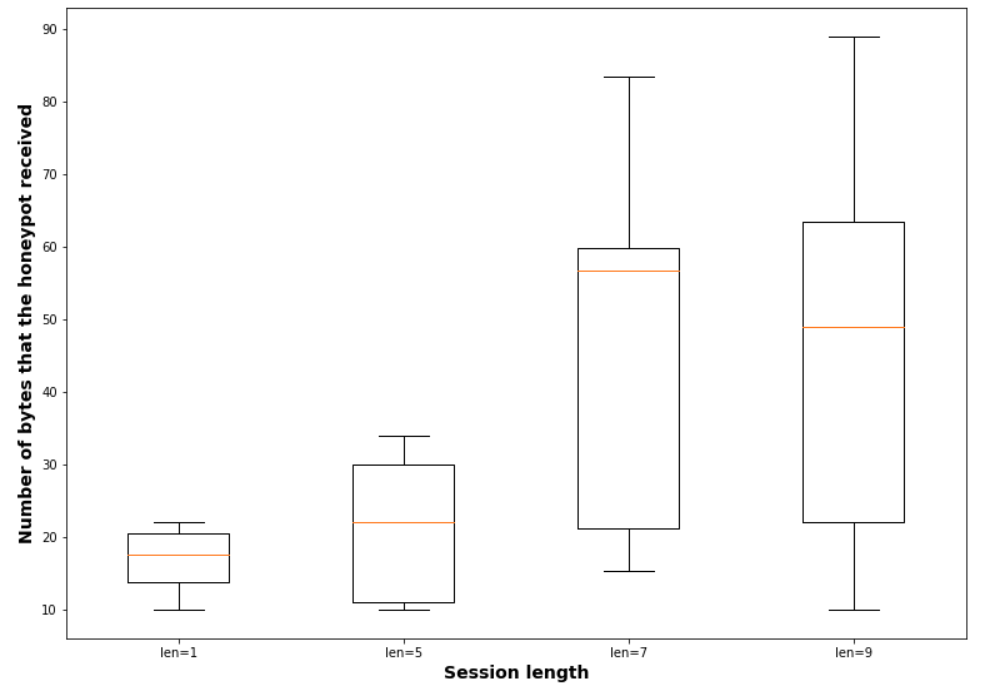}}
    \caption{Volume of the information that the honeypot receives per session length.}
    \label{fig:experiment2}
\end{figure}

Measuring the interaction length is not enough to understand whether this approach is more effective than others for understanding the attackers’ behaviours. The information collected from these interactions counts more than their length.
So, measuring the extraction of useful information from attackers is more important. We measured the volume of information sent by the attacker at a session time. Figure \ref{fig:experiment2} shows that the higher the interaction length is the greater the volume of the information sent by the attacker. This is reasonable because during a session time the more specific the request is the greater the size of the information sent.

\begin{figure}[htbp]
\centerline{\includegraphics[width=0.5\textwidth]{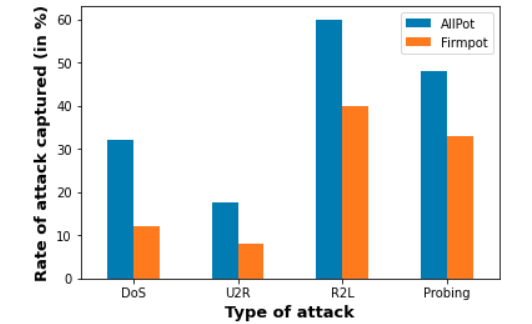}}
    \caption{Number of attacks captured by the honeypot based on NSL KDD dataset attack categories.}
    \label{fig:experiment3}
\end{figure}

Based on the classifier model we trained for the request evaluator module, we were able to evaluate the percentage of attack capture by our proposed honeypot. Figure \ref{fig:experiment3} shows that our honeypot captured about 30\% of Denial of Service (DoS) attacks 60\% of Remote to Locale (R2L) attacks which correspond to attacks such as password guessing for login attempts, query to the database, configuration changes, and more. User to Root (U2R) and probing attacks have also been captured. This gives us a great overview of the behaviour of attackers on our system.

\section{Conclusion}\label{sectionConclu}
Due to the large number and heterogeneous variety of IoT devices, building honeypots for IoT devices is very challenging when  using traditional methods. However, an attacker tends to perform preliminary checks on the device information before launching an attack. So, if a honeypot does not have a proper interaction mechanism with the attacker at the preliminary check stage, it is extremely hard to capture the complete exploit code. We have proposed an intelligent-interaction honeypot based on machine learning concepts to learn and interact with attackers automatically. Our evaluation indicates that the system can improve the session length with attackers and capture more attacks such as requests on the database, configuration changes, and login attempts. Our method helps us to collect more data for future training. However, we observe very few configuration change requests among the interactions with session lengths less than 7. This could be due to the fact that the honeypot has been discovered as a non-IoT device. It could also be that the attacker has corrupted our machine-learning model by poisoning  the dataset on which the model is trained to train. In practice, our model is running online and updated in real time. Therefore, it may not converge and reach the global optimal which corresponds to other reasons why we observe very few configuration change requests among the interactions with session lengths less than 7.

\section*{Acknowledgment}

This research was sponsored by the Army Research Office and was accomplished under Grant Number W911NF-21-1-0326. The views and conclusions contained in this document are those of the authors and should not be interpreted as representing the official policies, either expressed or implied, of the Army Research Office or the U.S. Government. The U.S. Government is authorized to reproduce and distribute reprints for Government purposes notwithstanding any copyright notation herein.

The work of the second\cmmnt{\sout{this} {\color{red}{the second}}} author is \cmmnt{{\color{red}{partly}}}partly supported by the EPSRC grant EP/V049038/1 and the Alan Turing Institute under the EPSRC grant EP/N510129/1.

The first author would like to thank\cmmnt{{\color{red}{The first author would like to thank}} \sout{Thanks to}} Moeka Yamamoto for providing the first part of the dataset used in the numerical \cmmnt{\sout{for the} {\color{red}{used in the numerical}}} experiments\cmmnt{{\color{red}{s}}} of this work, and\cmmnt{{\color{red}{and}}} also for the useful discussion in relation to the numerical work in this work.\cmmnt{ {\color{red}{in relation to the numerical work in this work.} \sout{for the evaluation of this work.}} }

\vspace{12pt}

\end{document}